\documentclass{ws-ijmpa}

\begin{document}

%

\def\nocropmarks{\vskip5pt\phantom{cropmarks}}

\let\trimmarks\nocropmarks      

%


%
%


\title{DOUBLE SPIN ASYMMETRY IN SEMI-INCLUSIVE\\
 DEEP INELASTIC SCATTERING\footnote{Talk given at the $3^{rd}$ Circum 
Pan-Pacific Symposium on High Energy Spin Physics, Beijing, China, Oct., 8-13,
2001}}

\author{\footnotesize ELVIO DI SALVO}

\address{Dipartimento di Fisica and I.N.F.N. - Sez. Genova, Via Dodecaneso, 
33 \\
16146 Genova, Italy\\}

\maketitle


\begin{abstract}
We shortly review the various methods suggested for determining the 
transversity function. Among such methods, we consider especially those 
based on semi-inclusive deep inelastic scattering. In the framework of this kind
of reactions, we propose to measure a double spin asymmetry, using a 
transversely polarized proton target and a longitudinally polarized lepton 
beam, and fixing the direction of the final pion. Under particular 
conditions, the asymmetry is sensitive to the transversity function.
\end{abstract}

\section{Introduction}
$~~~~$ The transversity function, $h_1$, may yield nontrivial information on 
the nucleon structure. However, such a distribution is quite difficult to 
determine experimentally. In fact, in the last years, the problem has been
debated at length\cite{ja01,ja1,ja2,ji,jj2,ba,dis,ef,ar}; 
moreover, the data analysis of the recent 
HERMES experiment\cite{her} has met serious difficulties. In this situation any 
observable sensitive to $h_1$ should be taken into account. The aim of the 
present talk is to illustrate one such observable, consisting of a double 
spin asymmetry in Semi-Inclusive Deep Inelastic Scattering (SIDIS), using a 
transversely 
polarized proton and a longitudinally polarized charged lepton. We shall also 
review the various methods proposed in the literature for determining $h_1$, 
referring in particular to SIDIS, which 
presents some advantages over the other kinds of reactions.

In sect. 2 we recall the definition of $h_1$ and illustrate the kind of 
information we may extract from this function. In sect. 3 we show the 
difficulties concerning the measurement of the transversity function. Moreover
we give a short review of the various methods suggested in the literature, 
referring, in particular, to the SIDIS single spin asymmetry measured in the 
HERMES experiment\cite{her}. As an alternative, in sect. 4, we examine the 
possibility of a SIDIS double spin asymmetry. We consider
two different cases, according as to whether the direction of the
final hadron is fixed or not. We treat in detail the former case, 
suggesting an alternative experiment for extracting $h_1$. 
Lastly in sect. 5 we present numerical estimates of the asymmetry illustrated 
in sect. 4 and give a short summary. 

\section{Definition of the transversity function}

The transversity function\cite{rs} is defined (see {\it e. g.}, Jaffe and 
Ji\cite{jj}) in 
terms of the tensor Dirac operator $\sigma_{\mu\nu}$. In order to understand 
the physical meaning of this distribution function, it is convenient to 
decompose the quark field into terms of given transversity. In a reference 
frame where the proton has a very large momentum directed perpendicularly 
to its polarization, it results that, for a given flavor $f$ and longitudinal 
fractional momentum $x$, $h^f_1 (x)$ is the difference between the number 
density of quarks with spin aligned along the proton spin and the number 
density of quarks with 
opposite spin. Using different projection operators\cite{jj}, we can establish 
two important properties of the transversity function, {\it i. e.}, that

(i) $h_1^f(x)$ is a twist-two distribution function, which amounts to saying 
that it survives in the scaling limit;

(ii) $h_1^f(x)$ is chiral-odd, which makes it difficult to determine 
experimentally this function, as we shall see in the next section.

This distribution is different from the helicity distribution $\Delta q^f(x)$,
for which one has to consider a proton travelling in the direction of its spin.
The reason is that generally a Lorentz boost does not commute with a rotation.
This would be the case, and the two distributions $h_1$ = $\sum_f e_f^2 h_1^f$ 
and $g_1$ = $\sum_f e_f^2 \Delta q^f$ (where $e_f$ is the fractional charge 
of the quark) would 
coincide, if the dynamics of the quarks inside the proton were 
non-relativistic. But we know that it is not so, because of the quark 
confinement 
and of the Heisenbrg principle; furthermore some predictions of the 
non-relativistic quark model fail, like the value of the axial charge. 
Therefore we may really expect nontrivial information on the nucleon structure 
from the determination of $h_1$. 
Indeed, some authors\cite{ss,ma} have stressed the importance of 
transverse momentum 
in the difference between $h_1$ and $g_1$. They have shown, in the framework of
the constituent quark model, that the quark tranverse momentum induces 
nontrivial 
Melosh-Wigner rotations, owing to the boost from the proton rest frame to the 
infinite momentum frame. This causes a spin dilution both in $g_1$ and in 
$h_1$. This dilution, in turn, may explain\cite{ma2}, at least partially, the 
so-called spin crisis, consisting of a surprisingly small value, found by the 
EMC collaboration in 1987\cite{emc}, of the first moment of $g_1$. But according
to the model the dilution is less marked in $h_1$ than in $g_1$. 
Therefore the determination of $h_1$, compared with $g_1$, 
may shed indirectly some light on the spin crisis. 

Furthermore, this determination could allow an important test for a QCD 
prediction on the $Q^2$ dependence of the two distribution functions. Indeed, 
while the QCD evolution of the singlet part of $g_1$ is coupled to the gluon 
polarization, which may produce sensible scaling violations, such an effect 
should be absent in $h_1$.
 
\section{Difficulties in determining $h_1$}

The difficulties in determining the transversity function are connected with 
its chiral-odd character. Indeed a massless quark conserves its chirality under 
any type of interactions, either electroweak or strong. It follows that, if a 
given asymmetry is sensitive to this function, it must depend on the product of 
$h_1$ by another chiral-odd function. Therefore we have to consider reactions 
in which two hadrons are involved, either in the initial or in the final state. 
Totally inclusive Deep Inelastic Scattering (DIS) is by no means suitable for
determining $h_1$. 

The double spin asymmetry in Drell-Yan (DY), with 
two transversely polarized proton beams, results to be 
proportional to\cite{ba} $\sum_f e_f^2 h^f_1(x_a) h^{\bar f}_1(x_b)$,
where $x_a$ and $x_b$ are the longitudinal fractional momenta of, respectively,
the active quark and antiquark that annihilate into a timelike photon. Here 
the drawback is that this asymmetry is quite small (1-2 \% at most\cite{ba}), 
moreover, presumably, $|h^{\bar f}_1|$ $<<$ $|h^f_1|$, since the antiquark 
necessarily belongs to the sea. Something we could gain using a polarized 
antiproton beam\cite{ba} instead of one of the two proton beams, but for the 
moment this kind of experiment looks quite unrealistic. 

More promising look asymmetry experiments based on SIDIS, that is, on 
reactions of the type 
\begin{equation}
\ell ~~ p \longrightarrow \ell' ~~ h ~~ X, \label{reaz}
\end{equation}
where $\ell$ is a charged lepton and $h$ a hadron. The proton target is 
polarized, which yields information on the polarization of the 
initial active quark. If also the lepton beam is polarized\cite{jj2}, 
or if the final hadron is a spinning, unstable particle, whose decay may be 
detected\cite{ma3}, we are faced with a double spin asymmetry. This allows to
get information on the final quark polarization and, in principle, 
to extract $h_1$. But even a single spin asymmetry, {\it i. e.}, with an 
unpolarized lepton beam, may be sensitive to the the final quark polarization,
provided we are able to exploit the so-called Collins effect\cite{coll} in the
angular distribution of $h$.

A single spin asymmetry is proportional to a 
mixed product of the type ${\bf S}\times {\bf p}_a\cdot {\bf p}_b$, where 
${\bf S}$ is the proton spin and ${\bf p}_a$ and ${\bf p}_b$ are any two (non 
collinear) momenta of the particles involved in the reaction. This object is 
invariant under parity inversion, but changes sign under time reversal. In 
other words, the cross section contains a $T$-odd term. This can only come from 
the interference between two amplitudes with different phases. In reaction 
(\ref{reaz}), this interference is produced by the final-state interaction 
between $h$ and other hadrons in the final state. Therefore in this
case it is the fragmentation function of $h$ that has a $T$-odd 
part, sensitive to the polarization of the fragmenting quark, as 
follows from the above mixed product. Here we identify ${\bf p}_a$ and 
${\bf p}_b$ with the momenta, respectively, of the final quark and of 
$h$, and approximate the quark momentum by the momentum ${\bf 
q}$ of the virtual photon. Then the problem amounts to determining the 
$T$-odd part of the fragmentation function.

To this end, consider the transverse momentum dependent (t.m.d.) fragmentation 
function of the final hadron, $\varphi^f (z, {\bf P}_{\perp})$, where $z$ and 
${\bf P}_{\perp}$ are, respectively, the longitudinal fractional momentum and 
the transverse momentum of the pion with respect to the fragmenting quark. 
While the usual 
fragmentation function, $D^f(z)$, is obtained simply by integrating $\varphi^f$
over the tranverse momentum, the $T$-odd part can be extracted by weighing 
$\varphi^f$ with the above mixed product, that is,
\begin{equation}
D^f_{odd} (z) = \int d^2 P_{\perp}\varphi^f (z, {\bf P}_{\perp}) sin\Phi, \ 
~~~~~~~ \ \ ~~~~~~~ \ sin\Phi = \frac{{\bf S}\times {\bf q}\cdot {\bf p}_h}
{|{\bf S}\times {\bf q}| |{\bf p}_h|}.
\end{equation}
Here ${\bf p}_h$ is the momentum of the hadron $h$. Moreover $\Phi$, the 
so-called Collins angle, is defined as the azimuthal angle between the (${\bf 
q}, {\bf S}$) plane and the (${\bf p}_h, {\bf S}$) plane. Notice that the 
$T$-odd fragmentation function is also chiral-odd. The Collins effect has been 
exploited in the recently realized HERMES experiment\cite{her}, where a 
longitudinally polarized 
proton has been used, and it is also invoked for the planned HERMES 
experiment\cite{he2} with a transversely polarized target. The asymmetries, 
respectively twist-3 and twist-2, are both sensitive to the product
$h_1^f(x) D^f_{odd}(z)$. Therefore the transversity function may be determined,
provided we are able to extract the Collins fragmentation function from an 
independent experiment. In any case, a  confirmation of the effect predicted by 
Collins comes both from HERMES data on the SIDIS single spin 
asymmetry\cite{her} and from $Z^0$ decay into two jets\cite{cern}. 

Variants of the Collins effect are the jet transversity determination\cite{ef} 
in DIS and the two-pion interference method\cite{ja1}, applicable both to 
SIDIS and to proton - proton collisions.

\section{Double spin asymmetry in SIDIS}

Alternatively we can, in principle, extract $h_1$ from a SIDIS double spin 
asymmetry experiment, employing a transversely polarized proton target and a 
longitudinally polarized lepton beam, and detecting a pion in the final state.
The asymmetry is defined as
\begin{equation}
A = \frac{d\sigma_{\uparrow\rightarrow}-d\sigma_{\uparrow\leftarrow}}
{d\sigma_{\uparrow\rightarrow}+d\sigma_{\uparrow\leftarrow}}, \label{as0}
\end{equation}
where the arrows denote the the proton and lepton polarizations. Two kinds of 
experiments are possible, that is, fixing the final pion direction, or 
integrating the cross section over the pion transverse momentum with respect to
the fragmenting quark. The latter possibility has 
been considered by Jaffe and Ji\cite{jj2} (JJ). The asymmetry they calculate 
contains twist 3 and twist 4 terms, the latters including the product
$h^f_1(x)\hat{e}^f_{\pi}(z)$; here $\hat{e}^f_{\pi}(z)$ is the twist-3 
fragmentation function of the pion, a chiral-odd function. Therefore, again, 
the determination of $h_1$ is subordinated to the knowledge, from an 
independent experiment, of another nonperturbative (and rather unusual) 
function. 

If we keep the final pion direction fixed, the asymmetry contains one more 
twist 3 term, which disappears upon integration over the pion transverse 
momentum. This term survives for ${\bf S}\cdot{\bf q}$ = 0, where the other 
terms - corresponding to the JJ asymmetry - vanish. Moreover, under this 
condition, such a term is sensitive to the 
t.m.d. transversity function, $\delta q^f (x, {\bf p}_{\perp})$. This can be 
intuitively seen by observing that, in this case, owing to the transverse 
momentum of the parton inside the hadron, the transverse polarization of the 
proton induces a longitudinal polarization in the active quark, which may be 
related to $\delta q^f (x, {\bf p}_{\perp})$. In principle there could be 
cancellations 
due to symmetries ({\it e. g.}, under rotation, parity inversion, etc.), but 
under the conditions we shall impose this does not occur. The situation is 
quite analogous to the one described in ref.\cite{dis}. There we considered a 
DY reaction of the type
\begin{equation}
p ~ p^{\uparrow} \longrightarrow \mu^+ {\vec \mu^-} ~ X,
\end{equation}
where we assumed to have one transversely polarized proton beam and to detect 
the longitudinal polarization of one final muon. In that case the asymmetry - 
which turns out to be the polarization of one of the muons - is non-zero, 
provided we consider nonvanishing, fixed values of the transverse momentum of 
the virtual photon with respect to the proton beams in the laboratory frame. 
Since a SIDIS reaction is kinematically isomorphic to DY, a similar effect 
occurs in the case considered, as we are going to show.

To this end we calculate the double spin asymmetry (\ref{as0}), taking 
${\bf S}\cdot{\bf q}$ = 0. In 
one-photon exchange approximation the differential cross section is of the type
\begin{equation}
d\sigma \propto L_{\mu\nu} H^{\mu\nu}, 
\label{dsg}
\end{equation}
where $L_{\mu\nu}$ and $H_{\mu\nu}$ are the leptonic and the 
hadronic tensor respectively. 
The leptonic tensor reads, in the massless approximation,
\begin{equation}
L_{\mu\nu} = k_{\mu} k'_{\nu} + k'_{\mu} k_{\nu} - g_{\mu\nu} k \cdot k' + 
i \lambda_{\ell}\varepsilon_{\alpha\mu\beta\nu} k^{\alpha} k^{'\beta}. 
\label{lept1}
\end{equation}
Here $k$ and $\lambda_{\ell}$ are respectively the four-momentum and the 
helicity of the initial lepton, $k'$ = $k-q$ the four-momentum of the final 
lepton and $q$ the four-momentum of the virtual photon. 

As regards the hadronic tensor, we use a QCD-improved parton model\cite{si}. 
The generalized factorization theorem\cite{qi1,bo2} in the covariant 
formalism\cite{land} yields, at zero order in the QCD coupling constant, 
\begin{equation}
H_{\mu\nu} \propto \sum_f e_f^2 \int d^2 p_{\perp} \sum_T 
q^{f}_{T} (x, {\bf p}_{\perp}) 
\varphi^{f} (z, {\bf P}^{2}_{\perp}) Tr (\rho^T \gamma_{\mu}\rho' 
\gamma_{\nu}).  \label{hadt}
\end{equation}
Here $q^f_T$ is the probability density function of finding a quark or an 
antiquark in a pure spin state, whose third component along the 
proton spin is $T$. Moreover the $\rho$'s are 
the spin density matrices of the initial and final active parton,
{\it i. e.},\cite{dis}
\begin{equation}
\rho^T = {1 \over 2} \rlap/p [1 + 2 T \gamma_5 (\eta_{\parallel} + 
\rlap/\eta_{\perp})], \ ~~~~~ \ ~~~ \rho' = {1 \over 2} \rlap/p'. 
\label{dens}
\end{equation}
$p$ and $p'$ = $p+q$ are, respectively, the four-momenta of the initial and 
final parton; moreover $2T\eta_{\parallel}$ is component of the parton 
polarization along its momentum and $2T\eta_{\perp}$ the quark transverse 
Pauli-Lubanski four-vector. $\eta_{\parallel}$ is a Lorentz scalar, such that 
$|\eta_{\parallel}|$ $\leq$ 1. It is immediate to check that eqs. (\ref{dens}) 
are consistent with the Politzer theorem\cite{po} in parton model 
apporoximation. Moreover  we have
\begin{equation}
{\bf P}_{\perp}  = {\bf \Pi}_{\perp}-z{\bf p}_{\perp},
\label{rel002}
\end{equation}
where ${\bf \Pi}_{\perp}$ is the transverse momentum of the pion with respect 
to the photon momentum. We keep ${\bf \Pi}_{\perp}$ fixed, therefore  
${\bf P}_{\perp}$ is a function of ${\bf p}_{\perp}$. Furthermore we
set $|{\bf \Pi}_{\perp}|$ $\leq$ 1 $GeV$, which is 
the condition for the factorization theorem to hold true\cite{coll,bo2}.

To calculate the asymmetry (\ref{as0}), we substitute eqs. (\ref{dsg}) to 
(\ref{dens}) into that expression, resulting in\cite{dis3} 
\begin{equation}
A(Q, x; y; z,{\bf \Pi}_{\perp}) = {\cal F}
\frac{\sum_{f=1}^3 e_f^2(\delta Q^f +\delta \bar{Q}^f)}
{\sum_{f=1}^3 e_f^2(Q^f +\bar{Q}^f)}, 
~~~~ \ ~~~~ {\cal F} = \frac{y(2-y)}{1+(1-y)^2}. \label{assidis} 
\end{equation} 
Here we have set $y = 1-E'/E$, where $E$ and $E'$ are, respectively, the 
initial and final energy of the lepton. Moreover we have introduced the 
quantities
\begin{eqnarray}
Q^f &=& \int d^2 p_{\perp} q^f(x, 
{\bf p}_{\perp}^2) \varphi^f (z, {\bf P}^{2}_{\perp}), 
\label{qf}
\\
\delta Q^f &=& 2 Q^{-1} \int d^2 p_{\perp} {\bf p}_{\perp}\cdot{\bf S} 
\delta q^f (x, {\bf p}_{\perp}) \varphi^f (z, {\bf P}^{2}_{\perp}), \label{dqf}
\end{eqnarray}
\begin{equation}
q^f  = \sum_{T = -1/2}^{1/2} q^f_T,  \ ~~~~~ \ ~~~~~ \ ~~~~~~ \
\delta q^f = \sum_{T = -1/2}^{1/2} 2T q^f_T \label{distr}.
\label{quef}
\end{equation}
$q^f$ is the t.m.d. unpolarized quark distribution.
We have slightly changed our notation, considering separately, for each 
flavor, the quark ($Q^f$, $\delta Q^f$) and antiquark ($\bar{Q}^f$, 
$\delta \bar{Q}^f$) contribution, the barred quantities being defined 
analogously to eqs. (\ref{qf}) and (\ref{dqf}). Some remarks are in order.

(i) Invariance of strong interactions under parity, time reversal and 
rotations (in particular rotations of $\pi$ around the proton momentum) implies
\begin{equation}
\delta q^f (x, {\bf p}_{\perp}) = \delta q^f (x, -{\bf p}_{\perp}).
\label{symde}
\end{equation}
This relation has two important consequences.
First of all the integral at the r. h. s. of eq. (\ref{dqf}) 
vanishes for ${\bf \Pi}_{\perp}$ = 0, therefore $\delta Q^f$ is proportional to
the the scalar product ${\bf \Pi}_{\perp}\cdot{\bf S}$. 
Secondly, if we consider totally inclusive DIS - which amounts to 
replacing $\varphi^f$ $\rightarrow$ 1 -, the integral (\ref{dqf}) 
is washed out by integration over transverse momentum. 

(ii) It is worth observing that, owing to the non-collinearity of the quark with
respect to the proton, the t.m.d. transversity function includes, unlike $h_1$, 
a chiral-even term, which 
can be calculated by changing the quantization axis from the proton 
momentum to the quark momentum. 
It is just such a chiral-even function that appears in formula 
(\ref{dqf}); this is why our asymmetry formula (\ref{assidis}), unlike the 
other asymmetries considered in the literature, does not contain any chiral-odd 
distribution or fragmentation functions. 

(iii) Gauge invariance implies that QCD first order corrections, in particular
graphs with one gluon exchange\cite{qi1}, contribute to the above mentioned 
asymmetry. However a calculation in the light cone gauge\cite{dis} assures that 
such contributions are about $10\%$ of the zero order terms.

(iv) Lastly the twist-3 character of the asymmetry (\ref{assidis}) - which can 
be immediately checked from eq. (\ref{dqf}) - forces us to pick up not too large 
values of $Q^2$ ($\leq$ 10 $GeV^2$). However this is not a serious limitation 
with respect to the twist-2 azimuthal asymmetries, which are plagued 
by a strong Sudakov suppression\cite{bo2} at large $Q^2$.

\section{Numerical results and summary}

Here we calculate the order of magnitude of the asymmetry (\ref{assidis}). To 
this end we assume\cite{bo2} $q^f$, $\delta q^f$ and $\varphi^f$ to have a 
gaussian transverse momentum dependence, with the same width parameter. Then
eq. (\ref{assidis}) results in
\begin{equation}
A(Q, x; y; z, {\bf \Pi}_{\perp}) = \frac{{\bf S}\cdot{\bf \Pi}_{\perp}}{Q} 
\frac{2z{\cal F}}{1+z^2}
\frac{\sum_f e^2_f\left[h_1^f(x)D^f(z)+\bar{h}_1^f(x)\bar{D}^f(z)\right]}
{\sum_f e^2_f\left[q^f(x)D^f(z)+\bar{q}^f(x)\bar{D}^f(z)\right]}.\label{formu}
\end{equation}
Taking into account eq. (\ref{formu}) and the second eq. (\ref{assidis}), we 
see that the optimal conditions for measuring the asymmetry are (i) $y$ and $z$ 
as close to 1 as possible and (ii) the pion transverse momentum relative to the 
photon parallel to the proton polarization. Under such conditions, and taking
$|{\bf \Pi}_{\perp}|\simeq 1$ $GeV$ and 
$Q$ = 2.5 $GeV$, we have $A \sim 0.4 R$, where 
$R = h_1^f(x)/q^f(x)$ has been determined by HERMES\cite{her}, $|R| = (50\pm 
30)\%$. 

To summarize, first of all, we have shortly reviewed the methods proposed in 
the literature for determining $h_1$. Then we have suggested a SIDIS 
experiment, with a 
longitudinally polarized lepton and a transversely polarized proton, detecting 
a pion in the final state. We demand to pick up events such that the lepton 
scattering plane is orthogonal to the proton polarization; moreover we select
pions produced in a fixed direction and at not too large angles with respect 
to the virtual photon momentum. The relative asymmetry is sensitive to the 
t.m.d. transversity function, but, unlike the other methods proposed in the 
literature, it does not involve any other unknown 
functions. The t.m.d. functions $q^f$ and $\varphi^f$ involved in asymmetry 
(\ref{assidis}) can be parametrized in a well determined way.
This asymmetry is estimated to be, for not too large
values of $Q^2$ and under favourable kinematic conditions, at least 
$\sim 10\%$. The experiment could be performed at some 
facilities, like CERN (COMPASS coll.), DESY (HERMES coll.) or Jefferson 
Laboratory, where similar asymmetry measurements have been realized or planned.

\end{document}